\documentstyle[12pt,amssymbols]{article}
\input math_macros.tex

\def\ref#1{$^{#1)}$}
\begin{document}
\begin{titlepage}
\begin{center}
\today     \hfill    LBL-31431 \\
          \hfill    UCB-PTH-59/91 \\
          \hfill    OHSTPY-HEP-T-92-003 \\
          \hfill    Stanford preprint ITP912/91
\vskip .25in

{\large \bf Predictions for Neutral $K$ and $B$ Meson Physics}
\footnote{This work was supported in part by the Director, Office of
Energy Research, Office of High Energy and Nuclear Physics, Division of
High Energy Physics of the U.S. Department of Energy under Contracts
DE-AC03-76SF00098 and DOE-ER-01545-573 and in part by the National Science
Foundation under grants PHY90-21139 and PHY86-12280.}

\vskip .25in
Savas Dimopoulos\\[.1in]

{\em Department of Physics\\
Stanford University\\
Stanford, CA 94305}
\vskip 9pt
Lawrence J. Hall\\[.1in]

{\em Department of Physics\\
     University of California\\
           and\\
      Theoretical Physics Group\\
      Lawrence Berkeley Laboratory\\
     1 Cyclotron Road\\
     Berkeley, California 94720}
\vskip 9pt
Stuart Raby\\[.1in]
{\em Department of Physics\\
The Ohio State University\\
Columbus, OH 43210}\\[9pt]
\end{center}

\vskip .2in

\begin{abstract}
Using supersymmetric grand unified theories, we have recently invented a
framework which allows
the prediction of three quark masses, two
of the parameters of the Kobayashi-Maskawa matrix and $\tan \beta$, the
ratio of the two electroweak vacuum expectation values.
These predictions are used to calculate $\epsilon$ and $\epsilon '$
in the kaon system, the mass mixing in the $B^0_d$ and $B^0_s$ systems,
and the size of CP asymmetries in the decays of neutral $B$ mesons to
explicit final states of given CP.

\end{abstract}
\end{titlepage}
\renewcommand{\thepage}{\roman{page}}
\setcounter{page}{2}
\mbox{ }

\vskip 1in

\begin{center}
{\bf Disclaimer}
\end{center}

\vskip .2in

\begin{scriptsize}
\begin{quotation}
This document was prepared as an account of work sponsored by the United
States Government.  Neither the United States Government nor any agency
thereof, nor The Regents of the University of California, nor any of their
employees, makes any warranty, express or implied, or assumes any legal
liability or responsibility for the accuracy, completeness, or usefulness
of any information, apparatus, product, or process disclosed, or represents
that its use would not infringe privately owned rights.  Reference herein
to any specific commercial products process, or service by its trade name,
trademark, manufacturer, or otherwise, does not necessarily constitute or
imply its endorsement, recommendation, or favoring by the United States
Government or any agency thereof, or The Regents of the University of
California.  The views and opinions of authors expressed herein do not
necessarily state or reflect those of the United States Government or any
agency thereof of The Regents of the University of California and shall
not be used for advertising or product endorsement purposes.
\end{quotation}
\end{scriptsize}

\vskip 2in

\begin{center}
\begin{small}
{\it Lawrence Berkeley Laboratory is an equal opportunity employer.}
\end{small}
\end{center}

\newpage
\renewcommand{\thepage}{\arabic{page}}
\setcounter{page}{1}

In previous papers [1] we have invented a predictive framework for quark
and lepton masses and mixings based on the Georgi-Jarlskog ansatz [2]
for the form of the mass matrices in supersymmetric grand unified theories [3].
In this paper we use this scheme to make predictions for parameters of the
neutral
K and B meson systems. In particular we show that the CP asymmetries in
neutral B meson decays are large and will provide a powerful test of the
scheme. We begin by reviewing the predictions for quark masses and mixings.

The top mass is predicted to be heavy
$$
m_t = 179 \:GeV \left({m_b\over 4.15 \:GeV}\right) \left({m_c\over 1.22
\:GeV}\right) \left({.053\over V_{cb}}\right)^2
\left( {1.46 \over \eta_b} \right) \left( {1.84 \over \eta_c} \right).\eqno(1)
$$
where $\eta_i$ is the QCD enhancement of a quark mass scaled from
$m_t$ to $m_i$.
Perturbativity of the top Yukawa coupling also requires that $m_t < $187
GeV.
In this paper the central values of $\eta_i$ quoted correspond to a complete
two loop
QCD calculation with $\alpha_s(M_Z) = .109$, whereas in reference 1 an
approximate two loop result was given.

The value of $\alpha_s =  .109$ comes from a 1 loop
analysis of the unification of gauge couplings which for simplicity
{\it ignored threshold
corrections at the grand unified and supersymmetry breaking scales} [1].
However these threshold corrections
will be present at some level in all grand unified models [4], and would not
have to be very large for our predicted value of the QCD coupling to range over
all values allowed by the LEP data: $.115\pm.008$. Whatever the threshold
corrections are, they must give an acceptable value for the strong coupling.
Hence it is important to consider the range of top masses allowed by the LEP
range of $\alpha_s$.
Larger values of $\alpha_s$ lead to larger $\eta_i$ reducing $m_t$.
Increasing $\alpha_s(M_Z)$ from .109 to .123 decreases $m_t$ from 179 GeV to
150 GeV. Alternatively, $\alpha_s(M_Z) = .123$ allows the top mass to be near
the fixed point value of 187 GeV with $V_{cb} = .047$. The above numbers refer
to the running mass parameter. The pole mass, which is to be compared with
experiment, is 4.5\% larger.

The particular form of the quark mass matrices leads to an unusual form
for the Kobayashi-Maskawa matrix
$$
V= \pmatrix{ c_1 - s_1s_2e^{-i\phi} & s_1+c_1s_2e^{-i\phi} &s_2s_3\cr
             -c_1s_2 -s_1e^{-i\phi} & c_1e^{-i\phi}-s_1s_2 & s_3\cr
               s_1s_3               &-c_1s_3               &e^{i\phi} }
\eqno(2)
$$
where $s_1= \sin \theta_1$, etc, and we have set $c_2=c_3=1$.
\footnote{The angle $\theta_3$ used in this paper corresponds to $\theta_3 -
\theta_4$ used in reference 1.}
We do not lose any generality by choosing the phases of quark fields such
that $\theta_1, \theta_2$ and $\theta_3$ all lie in the first quadrant.
We have predicted [1]
$$
\eqalignno{
s_1 &= .196\cr
s_2 &= .053 \chi &(3)\cr}
$$
where
$$
\chi = \sqrt{{m_u/m_d\over 0.6} \frac{1.22 GeV}{m_c} {\eta_c \over 1.84}}
\eqno(4)
$$
while the input $V_{cb}$ determines $s_3$ which must be chosen quite large in
view of (1).
The ratio  $m_s/m_d$, which we predict to be 25.15, strongly
prefers $m_u/m_d < .8$, while the present value of $|V_{ub}/V_{cb}|=
s_2$
prefers $m_u/m_d$ larger than .4.
In all of our predictions the largest uncertainty lies in $m_u/m_d$,
which we will display through the parameter $\chi$.

The angle $\phi$ is determined by the requirement that
$|V_{us}|=\sin\theta_c$
$$
\eqalignno{
c_\phi &= {1 \over \chi} \left( 0.51 \left( 1\pm 0.11 \right)
- 0.13 \chi^2 \right) = 0.38^{+.21}_{-.14}\cr
s_\phi &\simeq 0.92 \left( 1.15 - {0.15 \over \chi^2} \left( 1\pm 0.22 \right)
\right) = 0.92^{-.11}_{+.05}&(5)\cr}
$$
In the first expressions the $\chi$ dependence is shown explicitly,
together with the uncertainty from the measured value of the Cabibbo angle
$\sin\theta_c = 0.221 \pm .003$.
Note that the O($1\%$) uncertainties in $\theta_c$ become greatly magnified in
$\phi$. For this reason we keep track of the $\theta_c$ dependence in our
predictions.
For $\cos\phi$ the expression is exact,
while for $\sin\phi$ it is good to better than $1\%$.The final numerical
expressions correspond to the limits
$\chi^2 = 1\mp {1 \over 3}$ and $\sin \theta_c
= .221 \pm .003$, which are used for all numerical predictions in this paper.
Notice that $c_\phi$ is determined to be positive and the experimental
data on Re $ \epsilon$ in the kaon system forces $s_\phi$ positive.
Hence there is no quadrant ambiguity: choosing $\theta_{1,2,3}$ all in
the first quadrant means that $\phi$ is also in the first quadrant.
The rephase invariant measure of CP violation [5] is given in our model
by
$$
\eqalignno{
J&= Im V_{ud} V_{tb} V_{ub}^*V^*_{td} = c_1c_2c_3s_1s_2s_3^2s_\phi\cr
&= 2.6 \times 10^{-5} \left({V_{cb}\over .053}\right)^2 f(\chi)
&(6) \cr}
$$
where
$$
f(\chi)= \chi
\left( 1.15 - {0.15 \over \chi^2} \left( 1\pm0.22\right)\right)
=1^{-.29}_{+.23}.
\eqno(7)
$$
The scheme which leads to
these predictions
involves mass matrices at the unification
scale with seven unknown real parameters.
Six of these are needed to describe the eigenvalues: $m_u \ll
m_c \ll m_t$ and $m_d \ll m_s \ll m_b$, while the seventh is the CP violating
phase.
Hence a more predictive theory, having fewer than seven input
parameters, must either relate the up mass matrix to that of the down,
or must have an intrinsic understanding of the family mass hierarchy.
Without solving these problems the most predictive possible theory will
involve seven Yukawa parameters.
Such a predictive scheme can only be obtained by relating the
parameters of the lepton mass matrix to those of the down quark mass
matrix.
To our knowledge the only way of doing this while maintaining
predictivity is to use the ansatz invented by Georgi and Jarlskog [2].
The crucial point about our scheme for fermion masses is that
{\it it is the unique scheme which incorporates the GUT scale mass relations
$m_b = m_{\tau}, m_s = m_{\mu}/3$ and $m_d = 3m_e$ with seven or less Yukawa
parameters and completely independent up and down quark matrices.}
The factors of three result from there being three quark colors.

It is because of this uniqueness that the detailed confrontation of this
model with experiment is important.
If the model is excluded, for example by improving measurements of $m_t,
V_{cb}$ or $V_{ub}/V_{cb},$ then the whole approach of searching for a
maximally predictive grand unified scheme may well be incorrect.
Alternatively it may mean that there is a very predictive scheme, but it
involves relations between the up and down matrices in an important way.
It is important to
calculate the observable parameters of the neutral $K$ and $B$ meson
systems as accurately as possible, so that future experiments and
lattice gauge theory calculations will allow precision tests of this
scheme.

Our form for the Kobayashi-Maskawa matrix, equation (2), is very
unfamiliar and so we give predictions for the parameters that appear in
the Wolfenstein form of the matrix
[6].
The Wolfenstein form is an approximate form for the matrix which is
unitary only to order $\lambda^3
$, where $\lambda =\sin\theta_c =|V_{us}|$.
To order $\lambda^4$ we find the matrix can be written as:
$$
V^{(4)} =\pmatrix{ 1-{\lambda^2\over 2} - {\lambda^4\over 8}&\lambda&
A\lambda^4(\alpha -i\beta)\cr
-\lambda&1-{\lambda^2\over 2} - {\lambda^4\over 8} - {A^2\lambda^4\over
2} & A\lambda^2\cr
A\lambda^3 -A\lambda^4(\alpha +i\beta)& -A\lambda^2 +{A\lambda^4\over
2}& 1-{A^2\lambda^4\over 2}
}
$$
where $A$ is defined by $V_{cb}$ and $\alpha ,\beta$ by $V_{ub}$.
The reason that we prefer to work with the matrix at $0(\lambda^4)$
is that for us the $V_{ub}$ entry numerically really is $0(\lambda^4)$.
Thus we have $A, \alpha, \beta =0(1)$.
Notice that the order $\lambda^4$ contributions to $V_{ud}$, $V_{cs},
V_{ts}$ and $V_{tb}$ can be dropped unless an accuracy of greater than 2
1/2\% is required.
This means that $V^{(4)}$ is actually the same as $V^{(3)}$, the usual
Wolfenstein form with $\rho = \lambda\alpha$ and $\eta = \lambda\beta$.
Hence we use the usual Wolfenstein parameters $A,\rho, \eta$.
We find
$$
\eqalignno{
A     &= 1.09 {|V_{cb}|\over .053}\cr
\rho  &= {s_2c_\phi\over \lambda} = 0.12( 1-0.25 \chi^2) \cr
\eta  &= {s_2s_\phi\over \lambda} = 0.22 f(\chi). &(8)\cr}
$$

The leading dependence on the uncertain quantity $m_u/m_d$
is shown explicitly, through the parameter $\chi$.
The uncertainties in $\rho$ and $\eta$ coming from sin$\theta_c$
are less than $10\%$ and $4\%$ respectively. These are not shown as the
Wolfenstein approximation is itself only good to about $20\%$.
Notice that we can write $\rho + i\eta = re^{i\phi}$, where
$$
r = {s_2\over \lambda} = 0.24 \chi .\eqno(9)
$$
Requiring unitarity for the imaginary part of $V$ to $0(\lambda^5)$ [6]
one deduces
$$
J \simeq A^2 \lambda^6\eta = 3.1 \times 10^{-5} \left(
{V_{cb}\over .053}\right)^2 f(\chi). \eqno(10)
$$
This agrees well with the exact result of equations 5, since (10) is
expected to have $0(\lambda) \sim 0(20\%)$ corrections.

Beneath the scale of grand unification our effective theory is just that
of the  minimal supersymmetric standard model (MSSM).
Hence in the rest of this paper we wish to give the predictions for
$\epsilon, \epsilon ', x_d, x_s$ and the CP violating angles $\alpha,
\beta, \gamma$ in the minimal supersymmetric standard model, with the
Kobayashi Maskawa matrix given by equation 2, and with our predicted
values for quark masses.

In reference 7 it will be shown that in the MSSM the supersymmetric
contributions to $\epsilon, x_d, x_s$ and to the CP violating angles $\alpha,
\beta$, and $\gamma$ in B meson decay are small. Here we will simply give
a simplified
discussion of why these contributions are negligable for quark masses and
mixings of interest to us. In the standard model with a heavy top quark
the quantities $\epsilon, x_d$ and $x_s$ are dominated by box diagrams with two
internal top quarks. The amplitude of these standard model box diagrams can be
written as
$$
B_{ij}=A_{SM} \left( V_{ti} V^*_{tj} \right)^2\eqno(11)
$$
where $i,j=d,s,b$ label the relevant external
mass eigenstate quark flavors of the diagram and the
dependence on the Kobayashi-Maskawa matrix elements is shown explicitly.
The leading supersymmetric contributions to these three quantities come from
box diagrams with internal squarks and gluinos. In this case the flavor changes
occur through off-diagonal squark masses: $M^2_{ij}$. The amplitudes for these
box diagrams can be written as
$$
B'_{ij}=A_{MSSM} \left(M^2_{ij} \right)^2\eqno(12)
$$
where again only the relevant flavor structure has been shown explicitly.
It has been assumed that squarks of flavor $i$ and $j$ are degenerate.

In the MSSM the squarks are all taken to be degenerate at the grand unified
scale. The squark mass matrices evolve according to renormalization group
equations which generate non-degeneracies and flavor-changing entries. When all
quarks are light a very simple approximation for the flavor-changing entries of
the mass matrices results [8]. Since we predict a top Yukawa coupling close to
unity, this is not good enough for our purposes. We use the analytic solutions
of the renormalization group equations valid to one loop order in the top
Yukawa
coupling, but with other Yukawa couplings neglected [9]. This is the same
approximation used to obtain our quark mass and mixing predictions [1] and is
sufficient providing $\tan\beta$ is not so large as to make the bottom Yukawa
coupling large.
In this approximation only SU(2) doublet squarks have flavor changing masses.
We are able to find a very convenient approximation for the
induced flavor changing mass squared matrix elements for the down type
doublet squarks
$$
{M^2_{ij}\over M^2} \simeq 0.4 V_{ti} V^*_{tj}
\left(1+3\xi^2\over 1+5.5\xi^2 \right)\eqno(13)
$$
where M is the mass of the (nearly) degenerate squarks and $\xi$ is the ratio
of the gluino to squark mass at the grand unified scale. For values of the top
Yukawa coupling consistent with the prediction of equation 1, this
approximation is good to better than a factor of two. The exact result has only
slight sensitivity to the trilinear scalar coupling A of the MSSM, which we
have neglected.

Comparing the standard model box amplitude (equation 11) to that of the MSSM
superbox amplitude (equations 12 and 13) it is apparent that the dependence on
the Kobayashi-Maskawa matrix elements is identical. Hence the ratio of box
diagrams is independent of the external flavors i,j
$$
\left({B'_{ij}\over B_{ij}}\right)=
I\left(x\right) \left({100GeV\over M}\right)^2
\left(1+3\xi^2\over 1+5.5\xi^2\right)^2 \eqno(14)
$$
where $x=m_{\tilde{g}}/M$ and
$m_{\tilde{g}}$ is the gluino mass. The monotonic function
I results from the momentum integral of the superbox diagram
[8] and varies from I(1) = 1/30 to I(0) = 1/3. If the squarks are
taken light (eg 150 GeV) in an attempt to enhance the superbox amplitude,
then $x\geq 1$ to avoid an unacceptably light gluino,
resulting in $I \leq 1/30$.
To increase I therefore requires
an increase in M, but this rapidly decreases the importance of the superbox
diagram. The $\xi$ dependent factor in equation 14 is always less than unity.
We conclude that the supersymmetric contributions to $\epsilon, x_d$ and $x_s$
are unimportant in our scheme.

We have shown that, in the MSSM with degenerate squarks,
the superbox diagrams have the same
Kobayashi-Maskawa phases as in the standard model box diagrams. This implies
that the supersymmetric diagrams do not affect the CP asymmetry parameters
$\alpha,\beta$ and $\gamma$ in $B^o$ decay, as is well known [10].

We have assumed that at the grand unified scale the squark mass matrices are
proportional to the unit matrix. In supergravity theories this proportionality
is expected only at the Planck scale. In general it is possible that large
interactions of the quarks with superheavy fields could introduce large
flavor changing effects from renormalisation group
scaling between Planck and grand scales
[11]. We assume that this
does not happen, as would be the case if the only large Yukawa coupling in the
grand unified theory is that which generates the top quark mass.

We now proceed to our predictions.
Since the $K_L -K_S$ mass difference receives large long distance
contributions we do not think it provides a useful test of our theory.
On the other hand, all observed CP violation is described by the single
parameter $\epsilon$, which is reliably calculated from short distance
physics.
To calculate $\epsilon$ precisely we do not use the Wolfenstein form for
the KM matrix.
Instead we use a manifestly phase invariant formula for $\epsilon$ in
terms of $J$ [12].
We find that the box diagram with internal top quarks dominates.
Including a 20\% contribution from the diagrams with one top and one
charm we find
$$
|\epsilon |= 7.2 \ 10^{-3} B_K \left({ m_t\over 176 GeV}\right)^2
\left( {J\over 2.79 .10^{-5}}\right) \left({V_{cb}\over .053}\right)^2
{s^2_1\over s_c^2}.\eqno(15)
$$
The parameter $B_K$ describes the large uncertainty in the matrix
element of a four quark operator between kaon states. We use experiment for
$|\epsilon |$ and make a prediction for $B_K$:
$$
B_K =0.40 (1\pm.01 \pm.03)
\left({4.15 GeV\over m_b}\right)^2
\left({1.22 GeV\over m_c}\right)^2
\left(\eta_b \over 1.46 \right)^2
\left(\eta_c \over 1.84 \right)^2
f(\chi)^{-1}
\eqno(16)
$$
where we have used equations 1 and 6 for $m_t$ and J. The first uncertainty
shown comes from the experimental value of $|\epsilon|= (2.26\pm .02) \times
10^{-3}$ while the second comes from sin$\theta_c$.
This prediction for $B_K$ is strikingly successful.
We stress that $m_u/m_d$ cannot vary too much in our theory:
values larger than 0.6 are strongly disfavored by the fact that
$m_s/m_d$ is 25, while the present experimental values of $V_{ub}/V_{cb}$
disfavors $m_u/m_d$ lower than 0.6. Allowing $\chi^2=1\pm1/3$,
$\sin\theta_c=0.221\pm.003$ gives the range $B_K = 0.31 - 0.57$ (for
$m_b=4.15$ GeV and $m_c=1.22$ GeV). This should be compared with recent lattice
results $B_K = 0.7\pm0.2$ in the quenched approximation [13].

In the standard model, predictions for $\epsilon'/\epsilon$ are very
uncertain because they depend sensitively on:
i) various strong interaction matrix elements,
ii) the value of $\Lambda_{QCD}$,
iii) the value of the strange quark mass $m_s$ and
iv) the value of the top quark mass.
We use our central predictions for $\Lambda_{QCD}$ and $m_s$, and rely on the
1/N approximation for the QCD matrix elements [14]. Using the analytic
expression given in reference 14 we are able to derive our prediction
$$
{\epsilon'\over \epsilon} \simeq 3.9 \times 10^{-4}
\left(2.7 m_t^{-.5} + 0.5 m_t^{.1} - 2.2 m_t^{.4} \right)
\chi^{.4} \left( {0.4 \over B_K} \right)^{.8} \eqno(17)
$$
where $m_t$ is to be given in units of 179 GeV and is predicted in equation 1,
and $B_K$ is predicted in equation 16.
This small result is not unexpected, given the large value of $m_t$ [15], and
we
stress that it is uncertain because we do not know how well to trust the $1/N$
matrix elements.

The leading supersymmetric contribution to $\epsilon'/\epsilon$ comes from a
diagram with an internal gluino and an insertion of the flavor-changing squark
mass of equation 13. We find that this superpenguin amplitude is small compared
to the ordinary penguin:
$$
{A(superpenguin) \over A(penguin)} \simeq
0.04 \left( {150 GeV \over M} \right)^2
\left( {1+3 \xi^2 \over 1+ 5.5 \xi^2} \right)
\left( 5-4 \left({m_t \over 180 GeV} \right) ^2 \right)
^{-1} \eqno(18)
$$
for the case of degenerate squarks and gluino of mass $M$. This is partly
because the loop integral is numerically smaller, but is also because the
ordinary penguin diagram is enhanced by an order of magnitude by a large
$\ln m_t$ factor. Even though these supersymmetric contributions are
negligable, the uncertainties in the QCD matrix elements still imply that
$\epsilon'/\epsilon$ cannot be considered a precision test of our scheme.

The dominant
standard model contribution to $B^0\overline{B}^0$ mass mixing arises  from
the box diagram with internal top quarks.
We find that for the $B^0_d$:
$$
x_d = {\Delta m\over \Gamma} = .25 \left({\sqrt{B}f_B\over 150 \,MeV}\right)^2
\left({m_t\over GeV}\right)^2|V_{td}|^2.\eqno(19)
$$
Using equation (1) for $m_t, V_{td} = s_1s_3$ and the experimental value
for $x_d$ of $.67\pm .10$ we predict
$$
\sqrt{B} f_B = 134 MeV \left({4.15 GeV\over m_b}\right) \left({1.22
GeV\over m_c}\right)
\left(\eta_b \over 1.46 \right) \left(\eta_c \over 1.84 \right)
{|V_{cb}|\over .053}.\eqno(20)
$$
We note that if $x_d, m_t$ and $V_{td}$ are allowed to range over their
experimentally allowed values, the prediction of the box diagram
(equation 19) implies that $\sqrt{B} f_B$ will have to range over an
order of magnitude.
It is therefore a non-trivial success for our theory that it gives a
prediction for $\sqrt{B} f_B$ which is close to the  quoted
values for $f_B\sqrt{B}$. Our prediction should be compared with recent lattice
results: $f_B = 205\pm 40$ MeV and $\sqrt{B} f_B = 220\pm40$ MeV [16].
Our predictions for $m_t$ (1), $B_K$ (16) and $f_B$ (20) all depend
on $\eta_i$ which depend on $\alpha_s$. The numbers quoted are for the fairly
low value of $\alpha_s = .109$. Threshold corrections at the grand unified
scale could increase this, easily resulting in a 20\% increase in $\eta_b
\eta_c$. This not only reduces the top mass, but gives improved agreement with
lattice calculations for both $B_K$ and $f_B$. At any rate, once the top mass
is measured our predictions for $B_K$ and $f_B$ will be sharpened considerably.
Our results for $B_K$ and $f_B$ agree with those obtained reference 17.

The standard model box diagram relates the mass mixing in $B^0_s$ to
that in $B^0_d$ by:
$$
{x_s\over x_d} ={|V_{ts}|^2\over |V_{td}|^2}
\left({B_s f^2_{B_s}\over B_d f^2_{B_d}}\right)
= 25\left({B_s f^2_{B_s}\over B_d f^2_{B_d}}\right) \eqno(21)
$$
where we used our result for the ratio of Kobayashi-Maskawa
factors: $c^2_1/s^2_1$ = 25 with
negligible uncertainty.
If the ratio of $B$ meson decay constants could be accurately
calculated, and if large values of $x_s$ (say 15 to 25) could be
measured, then (21) provides a precision test of our theory.

Finally we consider CP asymmetries which result when $B^0$ and
$\overline{B}^0$ can decay to the same CP eigenstate $f$ [18].
Unitarity of the KM matrix implies the 1st and 3rd columns are orthogonal:
$V_{ud}V^*_{ub} + V_{cd}V_{cb}^* + V_{td}V^*_{tb}=0$.
This can be represented as a triangle since the sum of three vectors is
zero.
Labelling the angles opposite these three vectors as $\beta , \alpha$, and
$\gamma$ respectively, one finds that the CP asymmetries are proportional to
sin 2$\beta$ (for $B_d\to \psi K_s,$ etc), sin 2$\alpha$ (for $B_d \to
\pi^+\pi^-$ etc.) or sin 2$\gamma$ (for $B_s\to \phi K_s$, etc).

In the approximation that $c_2= c_3=1$
and that $s_1s_2 \ll 1$, we calculate sin 2$\alpha$, sin 2$\beta$ and
sin 2$\gamma$ to an
accuracy of (1 $+ 0(\lambda^3)$), ie to 1\% accuracy:
$$
\eqalignno{
\sin 2\alpha &= -2c_\phi s_\phi \cr
\sin 2\beta  &= {2c_1s_1s_2s_\phi\over s^2_c}
  ( 1+ {c_1s_2 c_\phi \over s_1})\cr
\sin 2\gamma &= 2c_\phi s_\phi {s^2_1\over s^2_c}
(1+ {c_1s_2  \over  c_\phi s_1}).
&(22)\cr}
$$
It is interesting to note that $s_3$ does not appear anywhere in these results.
This is because all lengths of the unitarity triangle are simply
proportional to $s_3$.
This is similar to the well known result that in the
Wolfenstein approximation all lengths of the triangle
 are proportional
to $A$.
For us this lack of sensitivity to $s_3$ is an essentially exact result.
Given that we know $s_1$ precisely,
and that $\phi$ is extracted from $s_1, s_2$ and the Cabibbo angle $s_c$, the
only uncertainties in numerically evaluating $\alpha, \beta $ and $\gamma$ come
from experimental uncertainties in $\sin\theta_c$ and in the dependence of
$s_2$
on $m_u/m_d$, $m_c$ and $\eta_c$ via $\chi$ shown in equations 3 and 4.
We calculate sin 2$\alpha$,
sin 2$\beta$ and sin 2$\gamma$ in terms of $\chi$ for sin$\theta_c = .221 \pm
.003.$
The results are shown in the Figure.
The solid line is for $\sin\theta_c =
.221$ while the long (short) dashed lines are for sin$\theta_c =
.224 (.218)$.
Present experiments allow
 very wide ranges of $\alpha, \beta, \gamma: \; -1 < \sin 2 \alpha ,
\sin 2\gamma < 1$ and $.1 <
 \sin 2\beta < 1$[19] so that our predictions are in a sufficiently narrow
range
that measurements of these CP asymmetries will provide
a precision test of our model.
Our predictions are very positive for experimentalists: $\sin 2\beta$ is not
near its lower bound, and for the two most experimentally challenging cases,
$\sin 2\alpha$ and $\sin 2\gamma$, the asymmetries are close to being maximal.

How well can our model be tested with an asymmetric $B$ factory operating at
the
$\Upsilon (4S)$ with luminosity of 3.10$^{33} cm^{-2}s^{-1}$ [20]?
We assume a total integrated luminosity of $10^{41}$ cm$^{-2}$, and find, using
the numbers in [19], that for decay to a final state of branching ratio $B$,
the quantity $\sin 2\alpha$ (or $\sin 2\beta$ or $\sin 2\gamma$) will be
measured
with an error bar $\pm\delta$:
$$
\delta = .05 \sqrt{ {B\over 4.10^{-5}}} \sqrt{ {10^{41} cm^{-2}\over \int
{\cal{L}}dt}}\eqno(23)
$$
where $B =(4,3,2)10^{-5}$ for $B_d \to \psi K_s$,
$B_d \to \pi^+\pi^-, B_s \to \rho K_s$ relevant for measuring sin 2$\beta$, sin
2$\alpha$ and sin 2$\gamma$ respectively.
Measuring all three quantities to $\pm .05$ will provide a spectacular
precision
test of our model.
The values of sin 2$\alpha$, sin 2$\beta$ and sin 2$\gamma$ must be fit by a
single value of $m_u/m_d$ which will be determined at the $\pm 0.1$ level.

We stress two important features of our predictions for these CP asymmetry
parameters. Firstly, as in the standard model, they are relatively insensitive
to unknown QCD matrix elements. Secondly, they test the Georgi-Jarlskog
ansatz in a deep way. For example the only dependence on the renormalization of
gauge couplings beneath the grand scale comes from uncertainties in $\eta_c$.
These uncertainties could be removed completely by taking the strange quark
mass as input. Taking $m_s = 180\pm60$ MeV only leads to a 15\% uncertainty in
$\chi$.

In this paper we have made accurate predictions for parameters in the neutral
$K$ and $B$ systems, in the belief that the scheme of reference (1)
will be decisively tested in the future.
It is worth stressing that the predictions for $B_K$ and $\sqrt{B} f_B$ are
close to central quoted theoretical values, and thus are already strikingly
successful.
We have followed the consequences of the only framework incorporating the
Georgi-Jarlskog mechanism which uses the minimal number of Yukawa couplings
and has independent up and down quark mass matrices:
there is
absolutely no guarantee that $\epsilon$ or $x_d$ will be predicted correctly.
Consider for example the case when $\epsilon$ is dominated by the top quark
contribution which is
proportional to $m_t^2 J B_K Re(V_{td}V_{ts}^*V_{us}V_{ud}^*)$.
Even though a theory may give successful predictions for $m_t (100 - 200 GeV)$,
$V_{td} (.003 - .018)$ and $V_{ts}(.030 - .054)$ it is not guaranteed that the
prediction for $\epsilon$ will be anywhere close to experiment.
The quantity $m^2_tRe(V_{td}V^*_{ts}V_{us}V_{ud}^*)$ has a spread of a factor
of 50, and the
quantity $J$ which is proportional to $s_1s_2s_3^2s_\phi$ could vary over a
very
wide range.
In particular recall that the phase $\phi$ is determined by the requirement
that
$|V_{us}|= .221 \pm .003$.
We think that it is extremely non-trivial
that the prediction of our theory for $m_t^2 Re (V_{td} V_{ts}^*
V_{us}V_{ud}^*)J$
is such that the central prediction for $B_K$ is 0.4.

The essential results of this paper are given in equation (16) for $B_K$ (from
$\epsilon$), equation (20) for $\sqrt{B} f_B$ (from $x_d$), equation (21)
for $x_s$ and in the Figure for sin 2$\alpha$, 2$\beta$, 2$\gamma$.
The prediction for $\epsilon'/\epsilon$ in equation (17) is less important as
it
involves uncertainties from the matrix elements.
Once the top quark mass is accurately known, the range of predicted values for
$B_K$ and $\sqrt{B}f_B$
will narrow.
The largest uncertainly in $B_K$ comes from $m_u/m_d$.
CP asymmetries in decays of neutral $B$ mesons offer the hope of a precision
test of our theory which is free of strong interaction uncertainties.
An asymmetric $B$ factory operating at the $\Upsilon$ (4$S$) with an integrated
luminosity of 10$^{41}$ cm$^{-2}$ can determine sin 2$\alpha, \beta, \gamma$
to an accuracy of $\pm 0.05$, and this will lead to a determination of
$m_u/m_d$
to within $\pm 0.1.$

\noindent{\bf Acknowledgements}

LJH acknowledges partial support from the NSF Presidential Young Investigator
Program, thanks Vernon Barger and Gian Giudice for discussions about QCD
corrections and thanks Uri Sarid for many helpful conversations.

\noindent{\bf References}
\begin{enumerate}
\item S. Dimopoulos, L.J. Hall and S. Raby, Phys. Rev. Lett. {\bf 68} 1984
(1992); LBL 31441, to be published in Phys. Rev. D.
\item H. Georgi and C. Jarlskog, Phys. Lett. {\bf 86B} 297 (1979).
\item S. Dimopoulos, S. Raby and F. Wilczek, Phys. Rev. {\bf D24} 1681 (1981).
S.Dimopoulos and H. Georgi, Nucl. Phys. {\bf B193} 150 (1981).
\item R. Barbieri and L. J. Hall, Phys. Rev. Lett. {\bf 68} 752 (1992).
\item C. Jarlskog, Phys. Rev. Lett. {\bf 55} 1039 (1985).
\item L. Wolfenstein, Phys. Rev. Lett. {\bf 51} 1945 (1983).
\item L. J. Hall and U. Sarid, in preparation.
\item M. Dugan, B. Grinstein and L.J. Hall, Nucl. Phys. {\bf B255} 413 (1985).
\item A. Bouquet, J. Kaplan and C. A. Savoy, Phys. Lett. {\bf 148B} 69 (1984);
Nucl. Phys. {\bf B262} 299 (1985).
\item C. Dib, D. London and Y. Nir, Workshop on Physics and Detector Issues
for a High Luminosity B Factory, Stanford,1990.
\item L.J. Hall, V.A. Kostelecky and S. Raby, Nucl. Phys. {\bf B267} 415
(1986).
\item I. Dunietz, Ann. Phys. {\bf 184} 350 (1988), and references therein.
\item S. Sharpe, Aspen Winter Physics Conference. Jan 1992.
\item G. Buchalla, A. Buras and M. Harlander, Nucl. Phys. {\bf B349} 1 (1991).
\item J. Flynn and L. Randall, Phys. Lett. {\bf B224} 221 (1989).
\item A. Abada et al. (European Lattice Coll.) I.N.F.N. Roma preprint 823.
\item V. Barger, M. Berger, T. Han and M. Zralek, MAD/PH/693(1992).
\item A. Carter and A. Sanda, Phys. Rev. Lett. {\bf 45} 952 (1980), Phys. Rev.
{\bf D23} 1567 (1981).
\item C. Dib, I. Dunietz, F. Gilman and Y. Nir, Phys. Rev. {\bf D41} 1522
(1990)
and references quested therein.
\item An Asymmetric $B$ Factory, Conceptual Design Report SLAC-372 (1991).

\end{enumerate}

\end{document}